\documentclass[pra,preprint,double-spaced]{revtex4}
\usepackage{amsmath}
\usepackage{graphicx}
\usepackage{latexsym}
\usepackage{amsfonts}
\usepackage{amssymb}
\usepackage{bm}
\usepackage{longtable}

\begin{document}
\title{Falsity of the Rouse Mode Solution of the Rouse and Zimm Models}
\author{George D. J. Phillies}
\email[To whom inquiries should be sent ] {phillies@wpi.edu}
\affiliation{Department of Physics, Worcester Polytechnic Institute,Worcester, MA 01609}

\begin{abstract}

The Rouse ({\em J. Chem.\ Phys.} {\bf 21}, 1272 (1953)) and Zimm ({\em J. Chem.\ Phys.} {\bf 24}, 269 (1956)) treatments of the dynamics of a polymer chain are shown to contain a fundamental mathematical error.  As a result, the oft-cited mode solutions for these models are qualitatively incorrect.  Comparison with the Wilson-Decius-Cross treatment of vibrational modes of polyatomic molecules reveals qualitatively the correct form for the solutions to the Rouse model.

\end{abstract}

\maketitle

For the past half-century, the  Rouse\cite{rouse} and Zimm\cite{zimm} models have provided a theoretical framework for the description of polymer dynamics in systems ranging from dilute solutions to the melt.  The objective of this note is to identify a critical mathematical defect in the Rouse mode solution for these models.  The nature of the failing is immediately evident from the Wilson-Decius-Cross\cite{wilson} treatment of small-displacement vibrations of isolated molecules. The failure, which is fundamental and arises at a mathematical rather than a physical level, has as a result that the characteristic modes and characteristic frequencies usually identified as solutions for these models are incorrect.  The following presents first the salient features of the Wilson-Decius-Cross and the Rouse-Zimm models, shows where their solutions contradict each other, and demonstrates where the Rouse-Zimm treatment went astray.

Wilson, Decius, and Cross\cite{wilson} treat small vibrations of an $N$-atom molecule.  For the displacements  ${\bf r}_{i}(t)$ of the atoms from their equilibrium locations, the equations of motion may be written
\begin{equation}
    m_{i}  \frac{d^{2} {\bf r}_{i}(t)}{d t^{2}}  = \sum_{j=1}^{N} {\bf V}_{ij} \cdot {\bf r}_{j}(t)
    \label{eq:WDCmotion}
\end{equation}
Here $m_{i}$ is the mass of atom $i$ and the ${\bf V}_{ij}$ are components of the  matrix of second derivatives (a $3N \times 3N$ matrix)  of the potential energy.  Eqn.\ \ref{eq:WDCmotion} represents a set of $3N$ coupled linear differential equations.   The solutions to these equations are a set of $3N$ eigenvectors (representing atomic displacements) and $3N$ corresponding eigenvalues $\lambda_{i}$, the $\lambda_{i}$ being squares of the corresponding vibration frequencies.  Six eigenvectors, each having $\lambda_{i} =0$,  correspond to whole-molecule translations and rotations that do not alter the relative positions of the atoms. The remaining $3N-6$ internal normal modes represent the molecular vibrations.

The degenerate zero eigenvalues create technical mathematical difficulties with solving the corresponding eigenvector-eigenvalue problem.  Wilson, et al.\cite{wilson} solved the problem by identifying an appropriate set of internal molecular coordinates, including bond stretches, bends, torsions, and out-of-plane motions, and a matrix method for transforming eqn.\ \ref{eq:WDCmotion} from its cartesian $3N$-dimensional form to a new $3N-6$-dimensional form. In the new form, the basis vectors are internal coordinates, while the whole-molecule translations and rotations occupy a 6-dimensional subspace orthogonal to the subspace of the internal coordinates. The Wilson-Decius-Cross FG method thus leads to a {\em non-singular} matrix whose solutions are the $3N-6$ internal modes.  One also implicitly obtains six eigenvectors having eigenvalue zero and corresponding to whole-molecule translations and rotations. For the simplest case of a bent triatomic molecule A-B-A, the $3N$ solution vectors are three orthogonal free translations, three orthogonal free rotations, and three internal modes qualitatively and imprecisely described as the bending mode, in which the A-B-A bond angle oscillates, and two bond stretching modes, in which the two A-B bond lengths oscillate in or out of phase with each other.

Comparison is now made with the Rouse and Zimm polymer models.  These models  treat a linear chain of frictional beads ("sub-molecules"), here labeled $(1, 2, \ldots ,N)$. The coordinate of bead $i$ is ${\bf r}_{i}$, the vector displacement from bead $i-1$ to bead $i$, $i=1$ being a special case. Because their origins move, the ${\bf r}_{i}$ form a set of non-inertial coordinates, but effects arising from the non-inertiality will be seen to be negligible.   Each bead has mass $m$ and drag coefficient $f$.  The distribution of bead-to-bead distances corresponds to a potential of average force whose gradients are the thermally averaged forces that beads exert on each other.  The distribution of distances is Gaussian, so the average force between adjoining beads is a harmonic restoring force having a spring constant here denoted $k$.  For a polymer chain in a stationary solvent, the Newtonian equations of motion corresponding to the Rouse model are (except for special-case beads $1$ and $N$) written
\begin{equation}
    m  \frac{d^{2} {\bf r}_{i}(t)}{d t^{2}}  =  - f \frac{d {\bf r}_{i}(t)}{d t} +  k( {\bf r}_{i+1}(t) + {\bf r}_{i-1}(t) - 2  {\bf r}_{i}(t))
    \label{eq:rousemotion}
\end{equation}
At low frequencies of interest here, the inertial terms forming the LHS of this equation, and matching fictitious forces arising from the non-inertial nature of the coordinates, are negligible. If the solvent is not stationary, additional forces $f {\bf u_{i}}$ appear, ${\bf u_{i}}$ being a local solvent velocity at the location of bead $i$. From symmetry, $\langle {\bf r}_{i} \rangle = 0$, so the ${\bf r}_{i}$ do represent displacements of the beads from their current equilibrium positions.

Several simplifications are now applied. The bead-bead average force represents a bond stretch. The angles between pairs of adjoining bonds do not affect the average forces. Beads $i$ and $i \pm n$, $n > 1$ have no direct interactions.  Therefore, the $x$, $y$, and $z$ displacements of the different atoms are seemingly uncoupled, allowing eq.\ \ref{eq:rousemotion} to be reduced to a set of $N-2$ equations for the $x$ coordinates of beads $2, 3, \ldots, N-1$,
\begin{equation}
   f \frac{d x_{i}(t)}{d t} =  k( x_{i+1}(t) + x_{i-1}(t) - 2  x_{i}(t)),
    \label{eq:rousemotion2}
\end{equation}
plus two-special-case equations for the $x$ coordinates of beads $1$ and $N$, and two corresponding sets of $N$ equations for the $y$ and $z$ coordinates.   The Zimm model adds to the forces on bead $i$ the hydrodynamic forces $\sum_{j} {\bf T}_{ij} \cdot {\bf F}_{j}$ due to mechanical forces ${\bf F}_{j}$ exerted on the solvent by the other beads $j$, ${\bf T}_{ij}$ being the pre-averaged Oseen tensor in which ${\bf F}_{j}$ and ${\bf T}_{ij} \cdot {\bf F}_{j}$ are parallel.

Eq.\ \ref{eq:rousemotion2} can be converted to a Langevin-type equation by adding a random thermal force ${\cal F}_{i}(t)$ on each bead.   Alternatively, after adding thermal forces $- k_{B}T \nabla_{i} \Psi$, Zimm converted eq.\ \ref{eq:rousemotion2} to an equation for the distribution function $\Psi({\bf r}_{1}, \ldots , {\bf r}_{N})$  of chain ends. These changes, while central to applications, are not relevant here.

Equations \ref{eq:rousemotion2} have for eigenvalues a single zero, corresponding to a mode in which all $N$ beads are displaced the same distance in the $x$ direction, and $N-1$ non-zero, non-degenerate relaxation times corresponding to $N-1$ internal modes for the $x_{i}$.  The relaxation times are
\begin{equation}
     \tau_{n} = \frac{f}{8k \sin^{2} (n \pi/2N)}
     \label{eq:rousetimes}
\end{equation}
for $n \in (1, 2, \ldots, N-1)$.  The displacements $x_{i}$ of the atoms are determined by the amplitudes $C_{n}$ of the normal modes via
\begin{equation}
      x_{i}  = C_{0}+ 2 \sum_{n=1}^{N-1} C_{n} \cos\left( \frac{\pi n (i-1/2) }{N}  \right),
      \label{eq:rousemodes}
\end{equation}
$C_{0}$ being the center-of-mass location, and with corresponding solutions for motions in the $y$ and $z$ directions. The relative  atomic displacements in a single mode $n$ are found by taking $C_{n}=1$ and all other $C_{j}$, $j \neq n$, to be zero.

The above coupled differential equations are separable, so to solve their spatial parts (the RHS of eqs.\ \ref{eq:WDCmotion} and  \ref{eq:rousemotion2}) it does not matter that their temporal parts (the LHS of the  eqs.\ \ref{eq:WDCmotion} and  \ref{eq:rousemotion2}) differ. Furthermore, from both a mathematical and a physical standpoint, the spatial parts of the Wilson-Decius-Cross and Zimm pictures, the RHS of eqs.\ \ref{eq:WDCmotion} and  \ref{eq:rousemotion2}, respectively, are the same. In each case, one has a set of coupled linear differential equations with constant coefficients.  In each case, the linear couplings are restoring forces that are linear in the distances between pairs of moving objects.  It is of no consequence that in one case the moving objects are called "atoms", and in the other the moving objects are called "beads".  It is of no consequence that in one case a coordinate is a displacement from an absolute equilibrium location, and in the other case a coordinate is a displacement from the current location of the prior bead in the chain.

The spatial parts of the Wilson-Decius-Cross and Rouse problems are thus identical in certain mathematical properties, so therefore the aspects of their solutions that are determined by those properties must also be the same.

Unfortunately, they are not.

The remainder of this paper exhibits for a simple model case the contradictions between the Wilson-Decius-Cross and Rouse solutions, and explains why the Rouse mode solutions are incorrect.

As a simple model, consider a bent triatomic molecule, such as water.  A single H-O-H molecule has six modes (three translations and three rotations) that do not displace the atoms with respect to each other, so that they  have $\lambda_{i} = 0$. The H-O-H molecule also has three internal vibrational modes with non-zero eigenvalues, namely a symmetric stretch mode, an antisymmetric stretch mode, and a bond bending mode.

Contrast the nine water modes with the modes of a three-bead Rouse model polymer, which may be obtained from eq.\ \ref{eq:rousemodes} by setting each of the $C_{j}$ seriatim to a non-zero value.  For the $x$ coordinates, the polymer has one mode of zero frequency, corresponding to uniform translation, one mode in which the first and third beads have equal and opposite displacements parallel to the $x$-axis while the center bead is stationary, and one mode in which the first and third beads move in the same direction along to $x$-axis while the center bead moves twice the distance in the opposite direction.  The modes for the $y$ and $z$ coordinates are the same (except for direction) as the $x$ modes.  The six stretch modes of the Rouse model are therefore trebly degenerate.   The Rouse model three-bead polymer has only three modes having eigenvalue zero, and six modes with non-zero relaxation times.

The disagreements between the Wilson-Decius-Cross and Rouse model normal modes are substantial.  Does a three-mass object with purely internal forces have six modes with eigenvalue zero, as predicted by Wilson, Decius, and Cross, or only three, as  predicted by the Rouse modes?  Do the internal modes have three distinct eigenvalues, or do they have two trebly degenerate eigenvalues? The answer is supplied by classical mechanics, which proves that purely internal forces that act along the lines of centers between masses within a body cannot affect the angular momentum of that body. The correct solution to the Rouse model, in addition to its three translational modes having $\lambda_{i} =0$, must also have three rotational modes not seen in eqs.\ \ref{eq:rousetimes} and \ref{eq:rousemodes} having $\lambda_{i} =0$.

Wherefrom comes the disagreement? Whole body rotations do not change bond lengths because they make simultaneous changes in bead positions along several axes.   To keep bond lengths constant during rotation, changes in bead locations along the $x$-axis are compensated by simultaneous changes in bond locations along the $y$ and $z$ axes.  The Rouse calculation breaks bead motions into three separate subspaces corresponding to the three coordinate axes, so that it has no  way to generate the coupled displacements in different directions that represent rotations without changes of bond length.

Mathematically, what has gone wrong is that the Rouse model solution applies standard eigenvalue-eigenvector methods to obtain the modes and relaxation rates of a singular matrix, namely the singular matrix $V_{ij}$ of coupling coefficients. The standard eigenvalue-eigenvector method gives invalid answers if the matrix of coefficients is singular, precisely as seen here.  The so-called Rouse modes, obtained by applying the standard eigenvalue-eigenvector method under circumstances such that the standard method is invalid, are not the actual modes of the Rouse model.

It might be proposed that the number of Rouse modes is actually extremely large, so even if only a few of them were incorrect the consequences might not be serious.  However, the zero-relaxation-rate modes that the Rouse modes fail to generate are the whole-body rotations.  Whole body rotations are the dominant modes for viscous dissipation and dielectric relaxation, and cannot be neglected.

It might be possible to apply some process analogous to the Wilson GF method to solve the Rouse model, at least for a chain in a particular given conformation. The potential energy matrix and set of restoring forces proposed by the Rouse model are purely and entirely a set of stretch restoring forces lying along the $N-1$ bonds connecting the $N$ beads of the chain.  There are no forces analogous to the Wilson-Decius-Cross three-atom and four-atom restoring forces that resist changes in bond angles, bond torsion angles, and bond out-of-plane displacements.  If there are no restoring forces that resist, e.g., changes in the angle between neighboring bonds, then modes that substantially act by changing those angles have eigenvalue zero.  For the three-bead model system, there is no restoring force in the bead-bead-bead bond angle, so the Rouse mode directly analogous to bond bending should have $\lambda_{i} = 0$. Should a GF-type solution route be attempted, the qualitative form of the solutions is immediately evident. The potential energy can be written in terms of precisely $N-1$ distance coordinates that determine all the non-zero contributions to the potential energy, so such a method would decompose the RHS of eq.\ \ref{eq:rousemotion2} into a $2N+1$ dimensional subspace having only zero eigenvectors, and an $N-1$-dimensional subspace whose $N-1$ modes are the true modes of the Rouse model.

It is unclear that a Wilson-Decius-Cross decomposition is the most profitable way to tackle this problem.  As an alternative, one might begin by calculating the contribution to polymer diffusion, viscosity, rotational diffusion, or dielectric relaxation of only six modes, namely the three translational modes and the three whole-body rotational modes.  The internal modes (necessarily orthogonal to the six chain modes being included) might then be treated as providing perturbations.  The original calculation of $\eta$ and $D_{s}$ by Kirkwood and Riseman\cite{kirkwood} implicitly uses this approach, while terming internal modes "fluctuations".  In this alternative approach, one remains in Cartesian coordinates, thereby permitting a straightforward analysis of hydrodynamic interactions between polymer coils, as seen in Refs. \onlinecite{phillies} and \onlinecite{merriam}.

To return to the primary conclusion, the standard mode solution to the Rouse model is incorrect.  The standard solutions fail to generate the whole-body chain rotations that make the dominant contributions to viscosity and other polymer solution properties . The terms are lost for fundamental mathematical reasons, not because the models are physically inadequate, with the consequence that calculations based on the mode solution may be invalid.

\end{document}